\documentclass[
showkeys,
twocolumn,
superscriptaddress,
showpacs,
amsmath,amssymb,
aps,
pre,
]{revtex4-2}
\usepackage{natbib}
\usepackage{graphicx}
\usepackage{dcolumn}
\usepackage{bm}

\begin{document}

\title{The emergence of Logarithmic-periodic oscillations in Contact
process with the topological disorder}
\author{Priyanka D. Bhoyar}
\affiliation{Department of Physics, 
S.K. Porwal College of Arts, Science and Commerce, Kamptee,441 001, India } 

\author{Prashant M. Gade }%
\email{prashant.m.gade@gmail.com}
\affiliation{
 Department of Physics, 
Rashtrasant Tukadoji Maharaj Nagpur 
University, Nagpur, 440 033, India.
}%
\date{\today}

\begin{abstract}

We present a model of contact process on Domany-Kinzel cellular automata
with a geometrical disorder.
In the  1-D model, each site is connected to two nearest neighbors
which are either on the left or the right.
The system is always attracted to an absorbing state
with algebraic decay of average density with a continuously
varying complex exponent. 
The log-periodic
oscillations are imposed over and above the usual power law
and are clearly evident as $p \rightarrow 1$.
This effect is purely due to an underlying topology
because all sites have the same infection probability $p$
and there is no disorder in the infection rate.
An extension of this model to two and three dimensions
leads to similar results.
This may be a common feature in systems where quenched disorder
leads to effective fragmentation of the lattice.
\end{abstract}


\maketitle
 {\it{Introduction: }}
The dynamic phase transition to a fully
absorbing vacuum state is the most studied
phase transition in
nonequilibrium statistical physics. 
Several universality classes have
been proposed for this transition. 
In simulations, the directed percolation
(DP) is the most observed universality class\cite{henkel2008non}. 
However, 
experimental verification of this universality class is obtained
in very few cases\cite{rupp2003critical,takeuchi2007directed}. 
The reasons could be the inevitable presence of noise and disorder in the experimental
systems\cite{vojta2006rare,webman1998dynamical}. 
The quenched disorder is a relevant perturbation if
the spatial correlation length critical exponent $\nu_\perp$
fulfills the condition
$\nu_\perp$\textit{d} $>$2 where \textit{d}
is dimensionality and $\nu_\perp$ is the correlation length exponent
in the spatial direction of the pure system \cite{harris1974effect}.
This is known as Harris criterion.

In some cases, quenched disorder leads to a whole parameter range
of very slow dynamics instead of a clean
critical point.
In this phase, the exponent of the power law is continuously changing
due to the formation of rare-region. This phase is known as Griffiths
phase \cite{griffiths1969nonanalytic,
vojta2003disorder}. 
This is in contrast with continuous phase transitions where
the power law associated
is observable only at the critical point. 

Power law in Griffiths phase has a real exponent usually. A complex
exponent will lead to log-periodic oscillatory corrections to the
power law \cite{sornette1998discrete}.
Complex  exponent has been obtained and studied in 
systems embedded with geometrical hierarchy, growth process,
rupture \cite{kapitulnik1983self,meurice1995evidence}.
They have also been identified in complex 
network\cite{odor2013slow,odor2013spectral}.
(This model does undergo fragmentation and underlying mechanism
may be similar to one proposed in this work.)
Recently, we have observed complex persistence exponent in 
a 1-d model where half of the sites obey  
rules leading to DP class 
and the rest evolve according to rules leading to compact directed 
percolation (CDP) class\cite{bhoyar}.
It demonstrates that such exponent may appear spontaneously
in systems without pre-existing hierarchy. Even for a random walk,
the discrete
scale invariance hierarchy is dynamically constructed
due to intermittent encounter with the slow region
\cite{bernasconi1983diffusion}.

In this work, we study the contact process on a d-dimensional lattice with
directed asymmetric coupling. We observe
log-periodic oscillations in the
decay of the fraction of infected sites $\rho(t)$.
This can be an outcome of
the quenched disorder leading to the effective
fragmentation of lattice. The model does not have a self-similar structure
in the defects or the lattice. 
This is a topological disorder.
Griffiths phase has been observed 
for complex networks which have inherent topological 
disorder\cite{munoz2010griffiths}.
The topological disorder may lead to the disappearance of
an active phase transition for the model of the resilience of
the internet against breakdown \cite{cohen2000resilience,cohen2001breakdown},
or disease spread for sufficiently
small infection rate\cite{wang2017spreading}. The model studied in this work
does not show an active
phase either. Throughout the phase diagram, we have an absorbing phase.
However, the dynamical approach to the vacuum state changes with the
parameter values.

Janssen-Grassberger conjecture 
\cite{grassberger1982phase,janssen1981nonequilibrium}
stated the conditions for DP transition. It can be stated as \cite{damage}
`the universality class of DP contains
all continuous
transitions from a “dead” or “absorbing” state to an “active” one
with a single
scalar order parameter, provided the dead state is not degenerate 
(and provided
some technical points are fulfilled: short range interactions both in
space and
time, nonvanishing probability for any active state to die locally, 
translational
invariance $[$absence of ‘frozen’ randomness$]$, and absence of multicritical 
points).'
We relax both of these conditions simultaneously.
In a 1-D model, every site is randomly labeled as R or L.
The site labeled R (L) is coupled to
two nearest neighbors on the right (left) side. 
This topological disorder results in
effectively partitioning the cluster into
several disconnected pieces as would be explained in the next section.
We extend the study in 2-D and
3-D and obtain similar results. 

\begin{figure}[hbt!]
\scalebox{0.3}{
\includegraphics{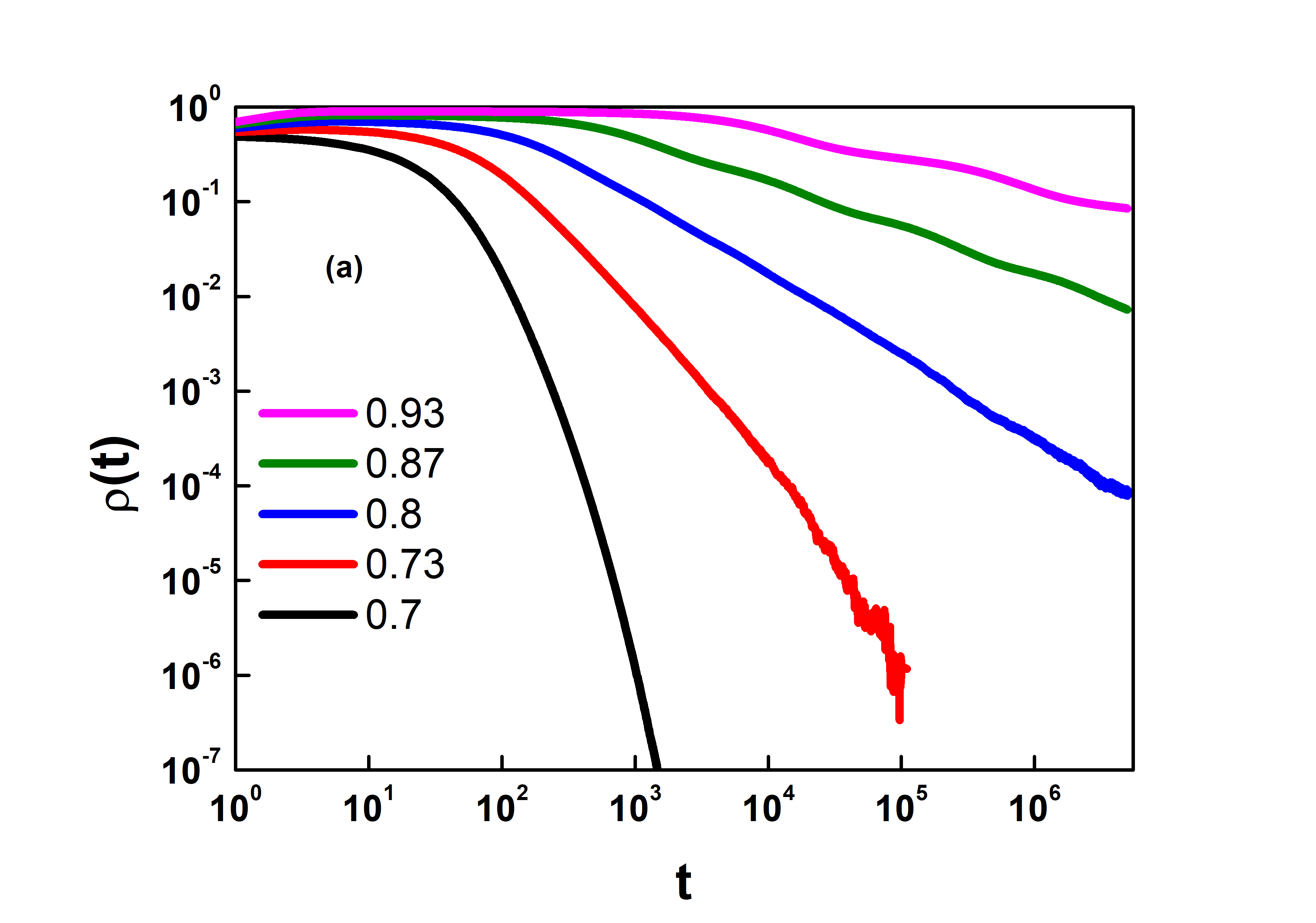}
}
\scalebox{0.3}{
\includegraphics{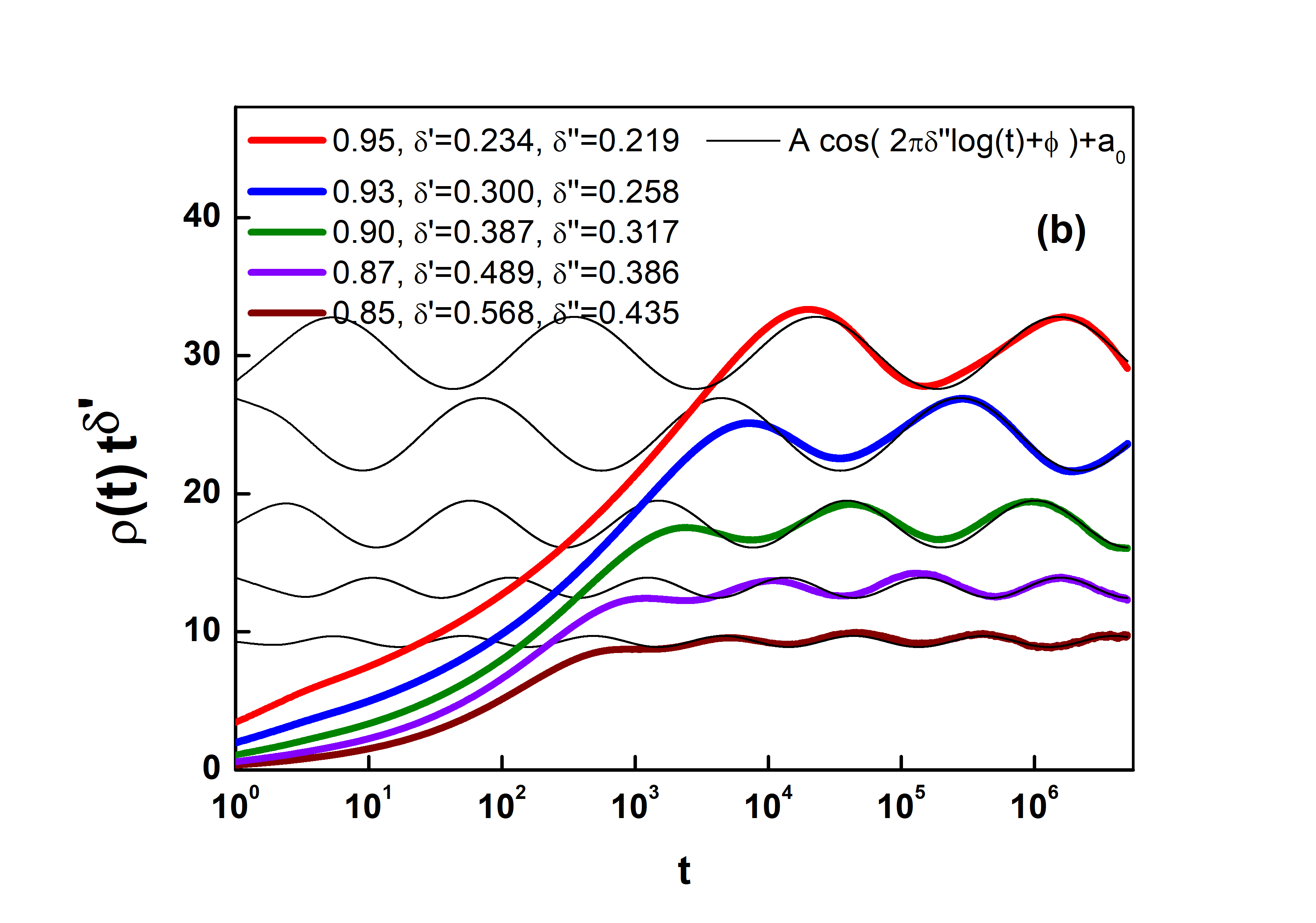}
}
\caption{(a) The time evolution of $\rho(t)$ $\it{vs}$ $t$ in 1-D model on log-log scale
, with $r=0.5$
and values of $p$  ranging for 0.7 to 0.93 (from bottom to top).
Clearly, the exponent of power law is changing continuously.
(b) We plot  $\rho(t) \times t^{\delta'}$ $\it{vs}$ $\log(t)$ for
 $p$ ranging from 0.85 to 0.95 (from bottom to top)
The log-periodic oscillations are clearly evident ($\omega=2\pi\delta''$. 
The y-axis is multiplied by an arbitrary constant for better visualization.}
\label{Fig:2}
\end{figure}

\begin{figure}[hbt!]
\scalebox{0.3}{
        \includegraphics{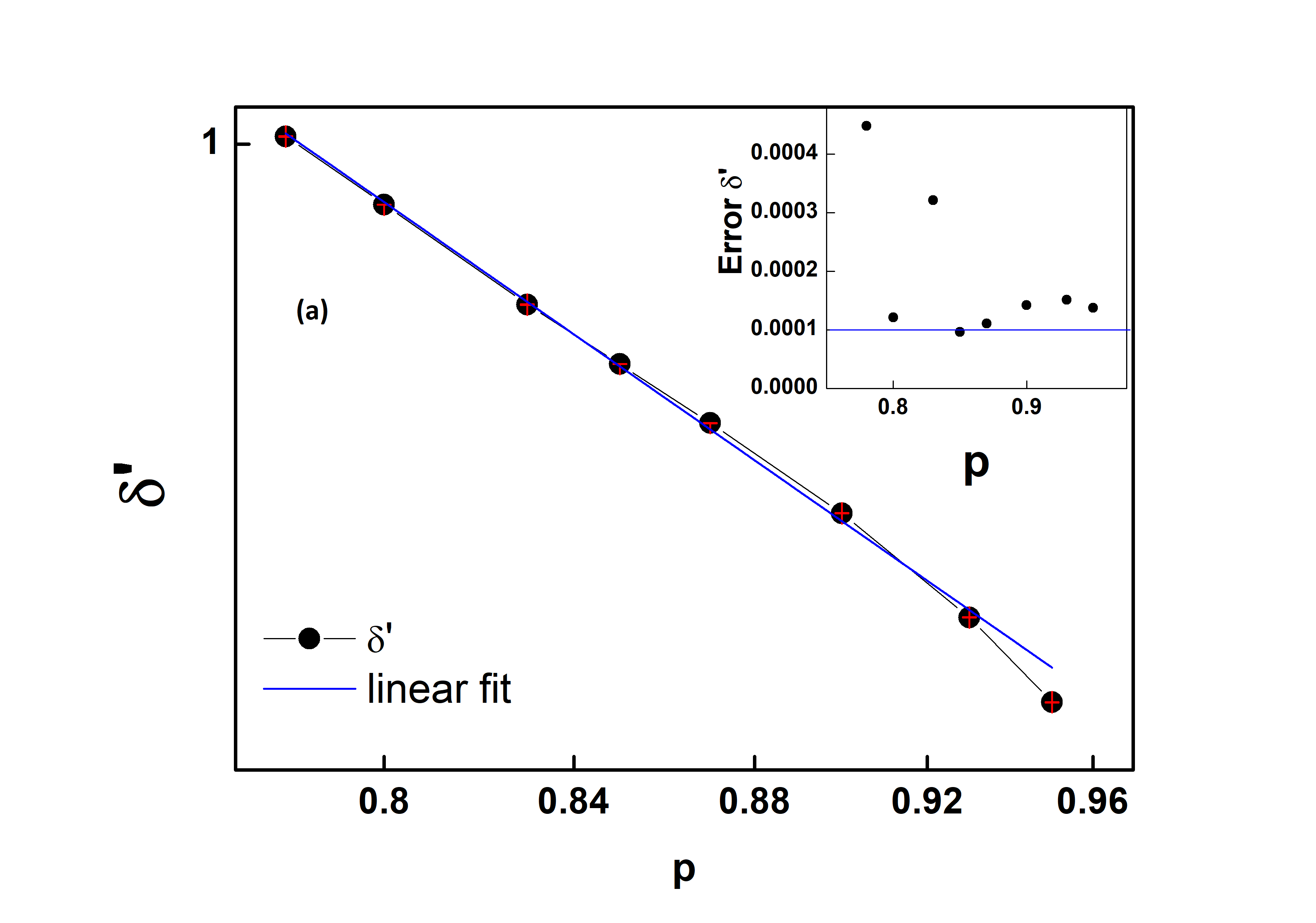}
}
\scalebox{0.3}{
        \includegraphics{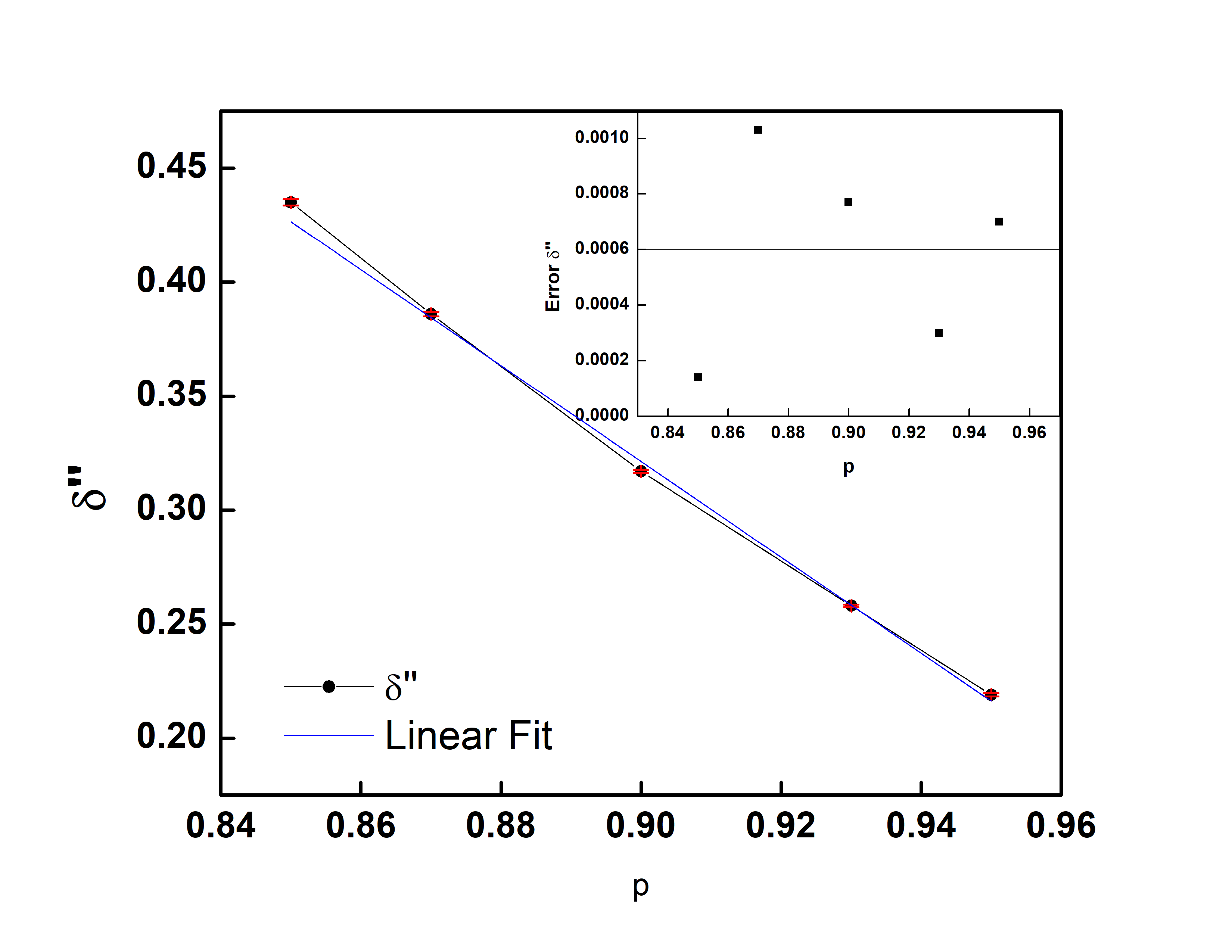}
}
\caption{(a) The log-log plot of $\rm{Re}(\delta)$ with various values of
$p$  ranging for 0.76 to 0.95. The error bar are also shown in figure(red marks).
(b)The log-log plot of $\rm{Img}(\delta)/2\pi$ with various values of
$p$  ranging for 0.83 to 0.95. The error bar are also shown in figure(red marks).
In both cases, the error in parameter is found using least
squares method and found to be smaller than  the size of symbols.
}
\label{Fig:3}
\end{figure}

{\it{Model and Simulation}}
We consider the cellular automata model of contact process  proposed
by Domany-Kinzel \cite{domany1984equivalence}.
The state of $i^\textit{th}$ site of a 1-D lattice ($\textit{v}_i=$0 or 1)
is specified at time t by $\textit{v}_i(t)$.
In the models of DP, sites marked as '1' could be interpreted
as wet or infected or chaotic while sites marked as '0' correspond
to dry or healthy or close to
the fixed point. 

\begin{figure}[hbt!]
\scalebox{0.3}{
        \includegraphics{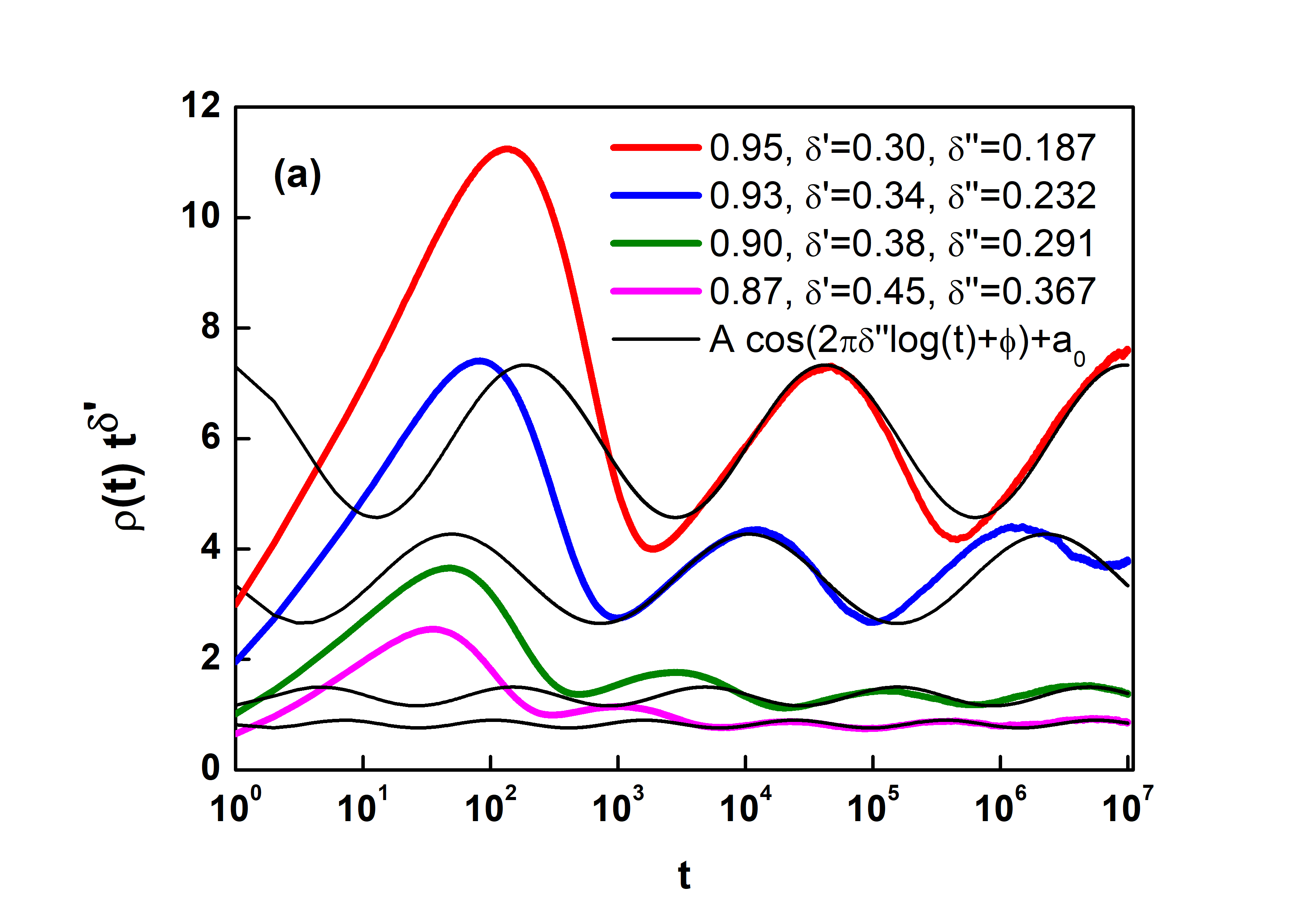}
}
\scalebox{0.3}{
        \includegraphics{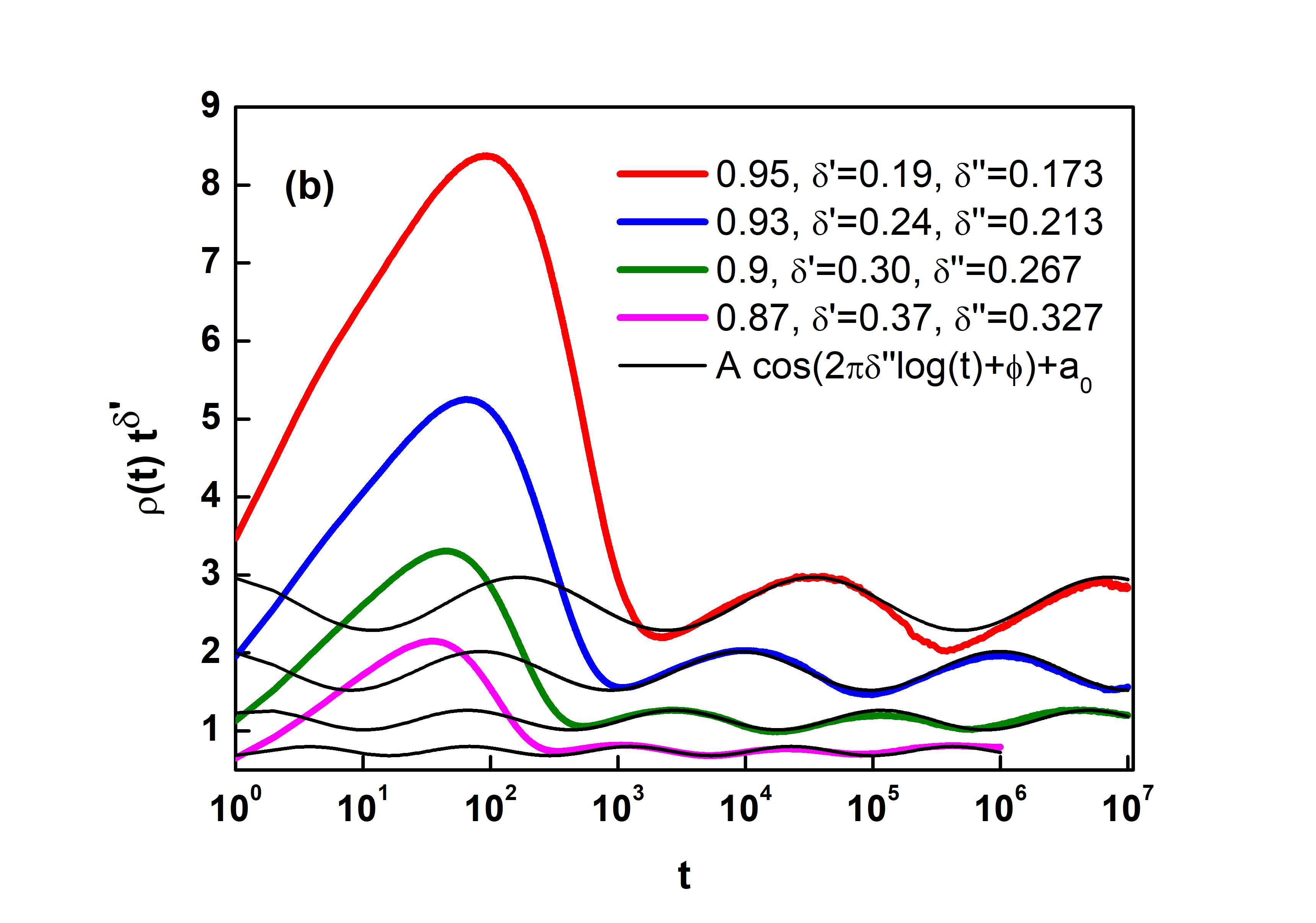}
}
\caption{(a) We plot time evolution of $\rho(t) \times t^{\delta'}$ for various
values
of $p$ in the range 0.87 to 0.95(bottom to top) in 2-D model.(b)
We plot $\rho(t) \times t^{\delta'}$ for values
of $p$ in the range 0.87 to 0.95(bottom to top)in 3-D model. 
The log-periodic oscillations are evident.
The y-axis is multiplied by an arbitrary constant for better visualization in both figures.}
\label{Fig:6}
\end{figure}

\begin{figure}[hbt!]
\scalebox{0.3}{
        \includegraphics{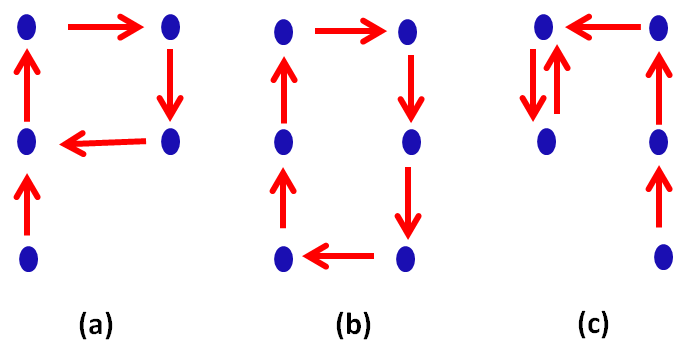}
}
\scalebox{0.3}{
        \includegraphics{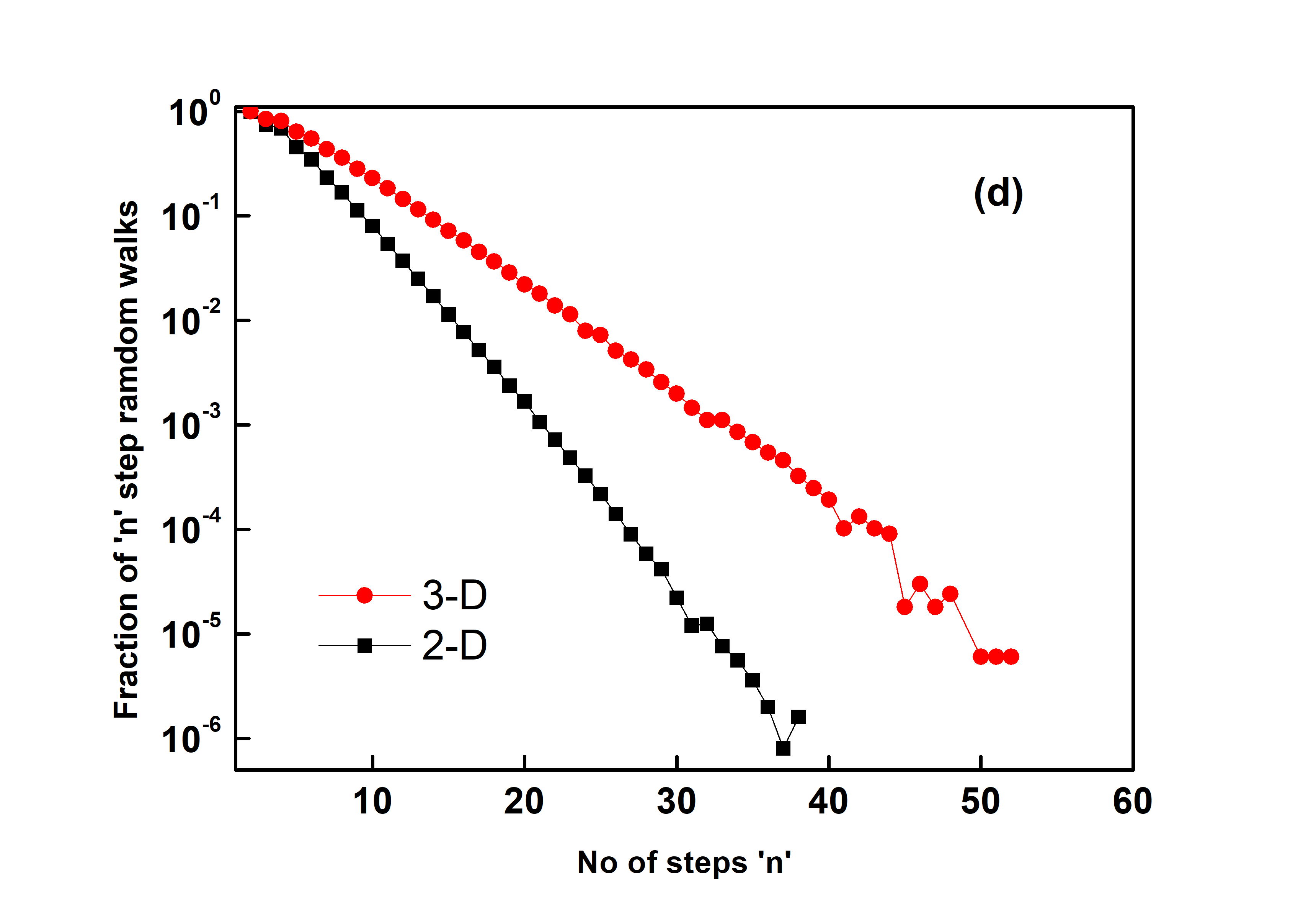}
}
\caption{(a),(b), and(c) A few examples of random walk in the 2-D model which ends up in an absorbing state. If the evolution of site A is affected by site B, we draw an arrow from site A to site B.
(d) shows the fraction of walks surviving till step $n$ in 2-D and 3-D.}
\label{Fig:7}
\end{figure}

Consider
a 1-D lattice of length N updated synchronously.
We introduce quenched disorder in the system 
by choosing a randomly chosen 
fraction $r$ of sites 
as type L and rest are chosen as type R.
We start with random initial condition with half of the sites chosen as 'active'. 
The sites L and R evolve according to conditional probability $P_L$ and
$P_R$. We define
$P_L(\textit{v}_i(t+1)|\textit{v}_{i-1}(t)+
\textit{v}_{i-2}(t))$  as
$P_L(1|0)=0$, $P_L(1|2)=P_L(1|1)=p$ and
$P_R(\textit{v}_i(t+1)|\textit{v}_{i+1}(t)+\textit{v}_{i+2}(t))$ is defined as
$P_R(1|0)=0$, $P_R(1|2)=P_L(1|1)=p$
where $p \ne 0$.The boundary condition are: if $i-1<0, i-2<0$,
then $i-1\equiv N+(i-1),i-2 \equiv N+(i-2)$
if $i+1>N, i+2>N$,then $i+1\equiv i+1-N, i+2 \equiv i+2-N$.
The order parameter, fraction of active sites is given by
$\rho(t)=\frac{1}{N}\sum_{i=1}^N v_i(t)$. The quantity $\rho(\infty)$
approaches zero asymptotically for $p<1$.
The updating scheme is synchronous.  It is noted for completeness.
 
We simulate the 1-D lattice of size
$N=2 \times 10^{6}$ for time up to $5 \times 10^{6}$ and
average over approximately $1000$ configurations.
We present the results for
$r=0.5$, {\it{i.e.}} half the sites are of type $L$ and the other half are of type $R$.
This is a case with the maximum disorder. At the same time, there is no anisotropy
on an average.
For small values of
infection probability $p$,
the fraction of active sites $\rho(t)$ undergoes exponential or stretched
exponential decay for small values of infection probability $p$.
(See Fig.\ref{Fig:2}(a).)
Nevertheless, as
$p\rightarrow 1$,
we observe a regime where 
$\rho(t)$ decays as a power law. The power law is given by
$\rho(t) \sim t^{-\delta}$, where the exponent is complex and
real part of $\delta$ is continuously
decreasing as $p \rightarrow 1$. The region of
continuously varying power law is known as the Griffiths
phase. Thus the above phase can be named as
a complex Griffiths phase.  The Griffiths phase 
usually
results due to the rare region effect. 

The Griffiths phase observed in the above connections has origin in 
effective fragmentation of lattice in disconnected parts of
different sizes.
Consider the sequence $RRLL$. The first two sites evolve according to
two sites on the right side and the next two sites evolve according to two
on the left. Thus, the evolution in these four sites
is practically independent of the rest of the lattice
(independent of the nature of update). Thus, if such a
group or any group which starts with $RR$ and ends
with $LL$  reaches
an absorbing state, it cannot come out of such a state.
We call such groups clusters of type 1.
Clusters will decompose
the entire lattice into several independent sections when
they reach absorbing state.
Now the decay
of the number of active sites will be dictated by sum of active sites
in several such independent
sections.  Consider any sequence which is
sandwitched between two consecutive type 1 clusters.
It will have $LL$ on the right and  $RR$ on the left.   This sequence
is essentially driven by type 1 cluster on the right side as well as 
left side. This sandwitched sequence of sites 
will
evolve independently of the rest of lattice
when clusters on either side become inactive. 
This group of sites is 
also a finite cluster. We call it type 2 cluster. Being
finite size, it will eventually 
reach an absorbing state when the clusters of type 1 on either
side become inactive. We have two types of clusters.
First is type one clusters which evolve 
essentially independently of  rest of lattice.
Type two clusters 
are sandwiched group of sites
between two type 1  clusters.
When both 
of them
reach an absorbing state, this cluster 
will eventually reach an
absorbing state.
These sections of type 1 and type 2 have
different lengths and the expected time by which they
reach an absorbing state is different as well. The absence of a
single length-scale (or time-scale) in evolution could lead to
non-exponential relaxation. We indeed obtain power law relaxation
over a large range of parameters in this model.

The lattice decomposes into several finite size clusters.
The activity of these
some large rare clusters is disassociated from the bulk
$\it{i.e.}$ although the bulk lattice is in the absorbing phase,
the 
rare regions are locally in the fluctuating phase. This leads to slow dynamics
in the Griffiths phase \cite{vojta2006rare}.
Due to the finite size of these disconnected sets of sites, the system
always collapses to the absorbing phase.(see Fig.\ref{Fig:2}(a)) for any
value of $p<1$. Thus, the model has only an absorbing phase 
and no fluctuating phase.

With no effective fragmentation of lattice,
DP universality class is restored even with 
quenched disorder in topology. 
With probability $r$, we connect a given site $i$ to  both nearest
neighbors and with probability $1-r$ we connect them
to both next-nearest neighbors.
The system undergoes DP transition with a 
clean critical point. 

The exponent of power law
is complex in nature.
A complex exponent can be written as $\delta=\delta'+i 2 \pi \delta''$.
Thus 
$\rho(t)\sim$ $Re(At^{-\delta'-i\delta''})$ $\sim$ $At^{-\delta'}
\cos(2 \pi \delta''\log(t))$ and 
$\rho(t)t^{\delta'}\sim A \cos(2 \pi \delta''\log(t))$.
As the function is log-periodic, it is very difficult to extract
the exact periodicity. The amplitude of these oscillation increases as
$p\rightarrow 1$.
These oscillations become more evident if we plot 
the quantity $\rho(t) \times t^{\delta'}$ 
with time
 Fig.\ref{Fig:2}(b). This behavior can be fitted by a constant superposed
by log-periodic oscillations. The amplitude and wavelength of these oscillations
grow as $p \rightarrow 1$.
For small values of $p$ the amplitude is very small(if any) and
it makes it difficult to determine if $\delta'' \ne 0$.
The value of $\delta'$
decreases as $p \rightarrow 1$ (See Fig.\ref{Fig:2}(b)).
Fig.\ref{Fig:3}(a) and Fig.\ref{Fig:3}(b) shows the linear fit of $\rm{Re}(\delta)$
and $\rm{Img}(\delta)/(2\pi)$
on log-log scale. The errorbar are shown as well. The error in case of $\delta'$ is calculated
from the goodness of linear fit of $\rho(t)$ vs t for various values of $p$. 
In case of $\delta''$, we use this value of $\delta'$ and
plot $\rho(t)t^{\delta'}$ as a function of $t$.
We have fitted a fuction 
$A\cos(2\pi\delta''\log(t)+\phi)+a_0$ 
and found error in $\delta''$
using nonlinear
least squares. 
A 'fit' function in gnuplot which 
uses an implementation of the nonlinear least-squares 
(NLLS) Marquardt-Levenberg algorithm was used\cite{ranganathan2004levenberg}.
This procedure has been followed in
for all fits to complex exponents in this work.
(Of course, the fit is carried out over the relevant range.
For larger values of $p$, the fragmentation of lattice occurs late
and the onset of logarithmic oscillations is delayed.)
The error in $\delta''$ is found to be very small
and less than $1\%$ in all cases. It is  less than $0.1\%$ for 1-d case.

\begin{figure}[hbt!]
\scalebox{0.3}{
\includegraphics{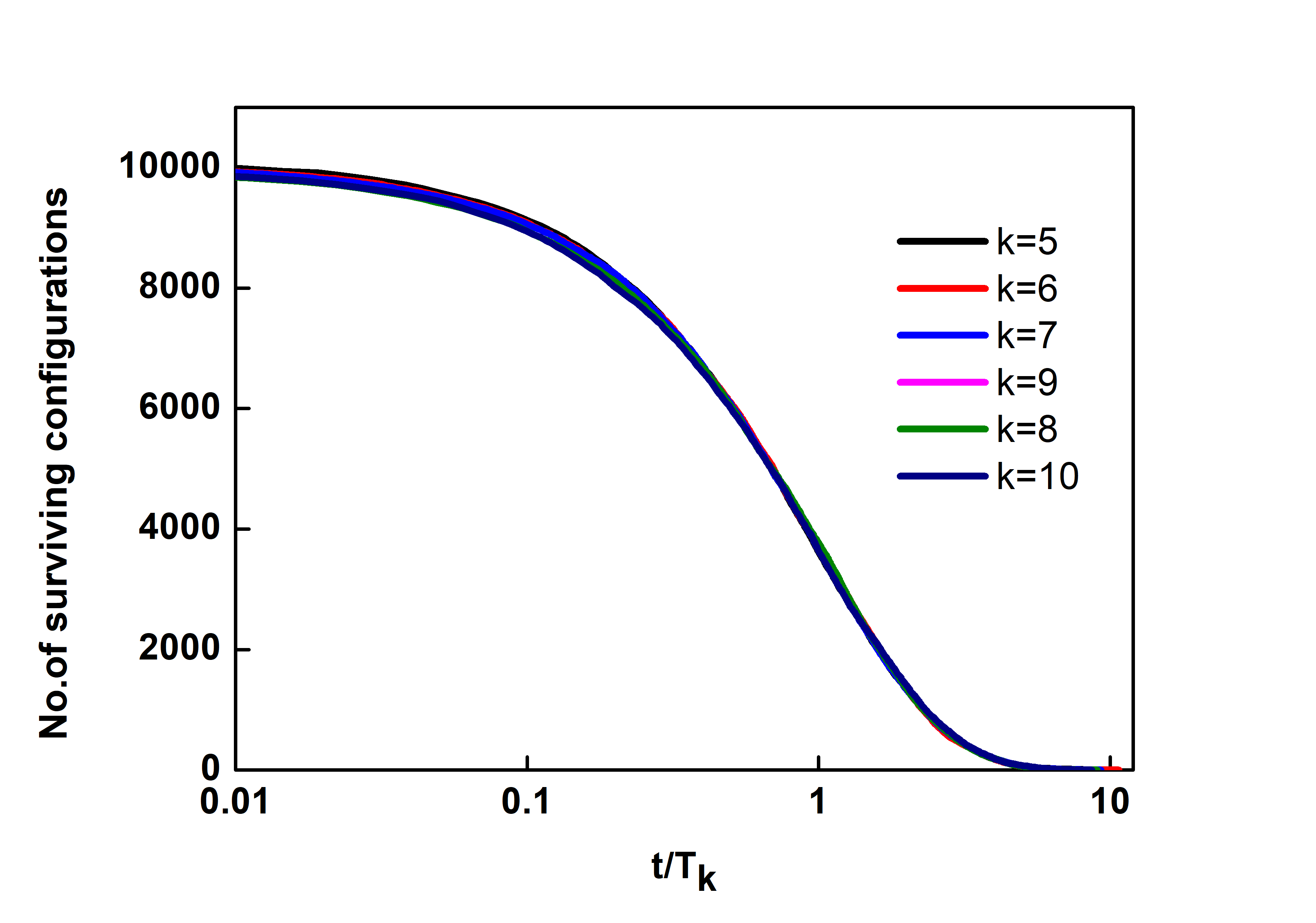}
}
\caption{We plot the number of surviving configuration $\it{vs}$
$t/T_k$ for  group of size $k=5,6,7,8,9,10.$ The initial number of
configurations are $10^4$ and $p = 0.94$.}
\label{Fig:4}
\end{figure}

\begin{figure}[hbt!]
\scalebox{0.3}{
\includegraphics{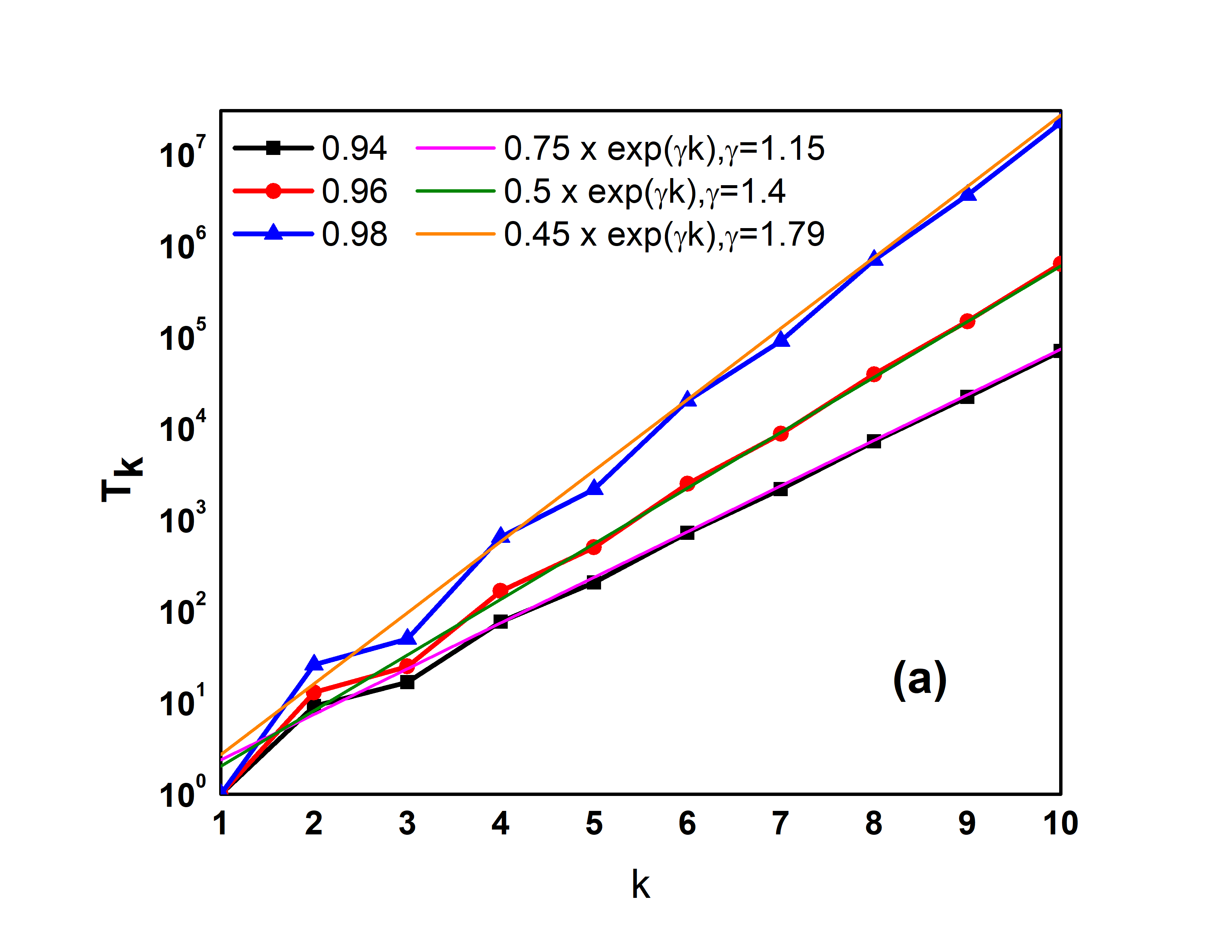}
}
\scalebox{0.3}{
\includegraphics{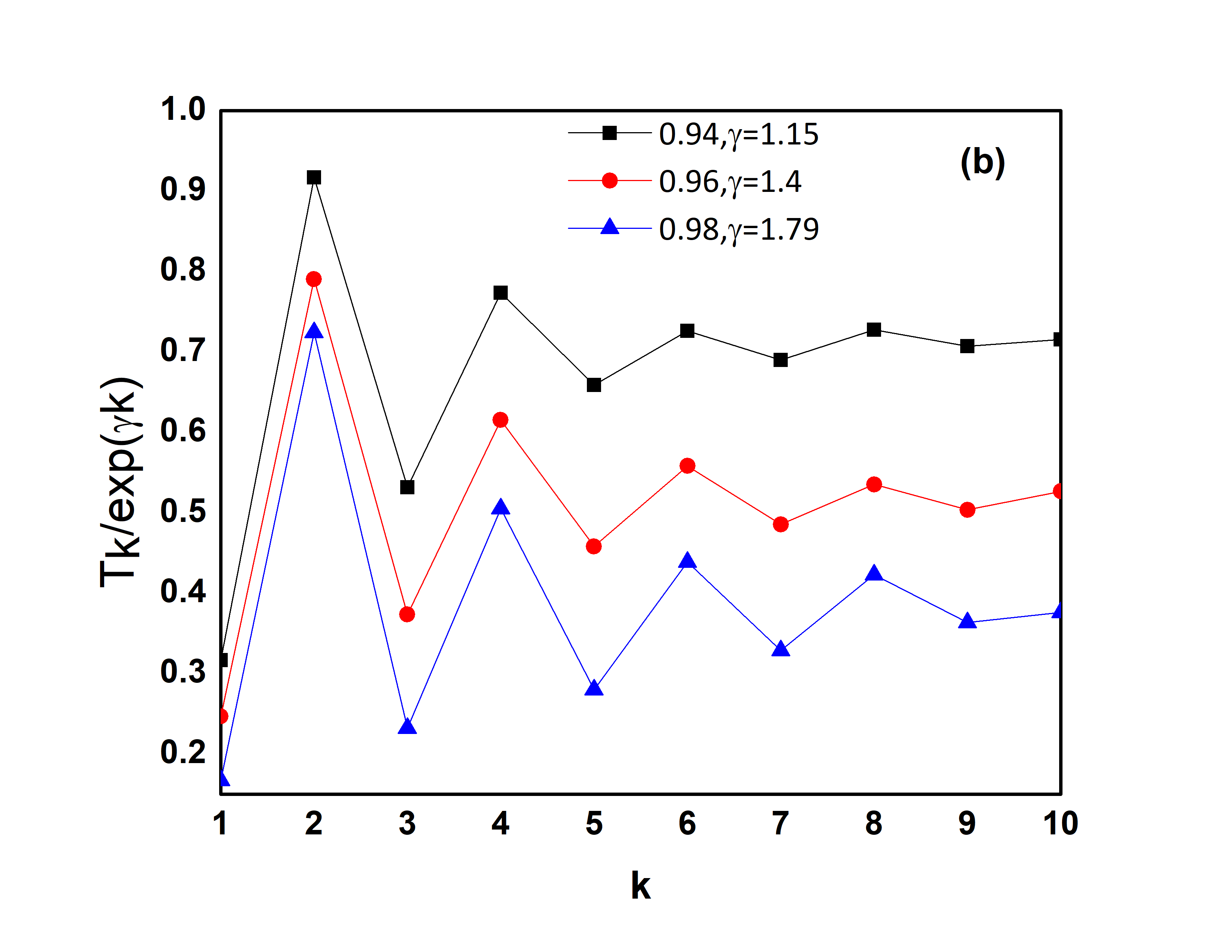}
}
\caption{(a)We plot $T_k \textit{vs}$ $k$, where $T_k$ is the average time taken by
cluster of k sites to become inactive starting with $10^4$ configurations. We
consider $k=1-10$. $T_k$ can be fitted as 
$\exp(\gamma_p k)$ where
$\gamma_p=1.15,1.4,1.79$ for $p=0.94,0.96,0.98$.
(b) We plot the relation $T_k/\exp(\gamma_p k) $ with $k$ 
for$p=0.94,0.96,0.98$.
The oscillations are pronounced for larger values of $p$.}
\label{Fig:5}
\end{figure}

\begin{figure}
\scalebox{0.3}{
\includegraphics{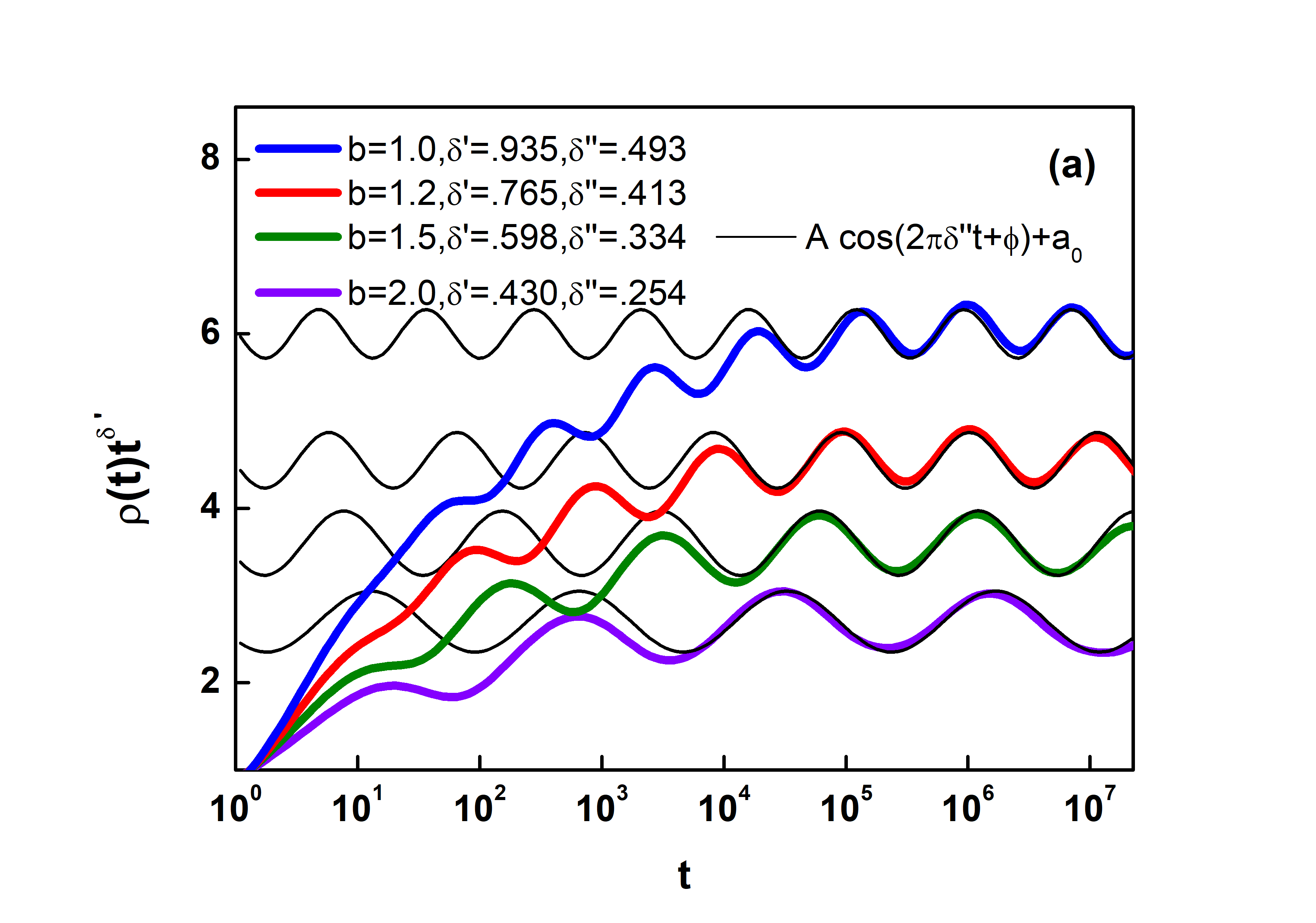}
}
\scalebox{0.3}{
\includegraphics{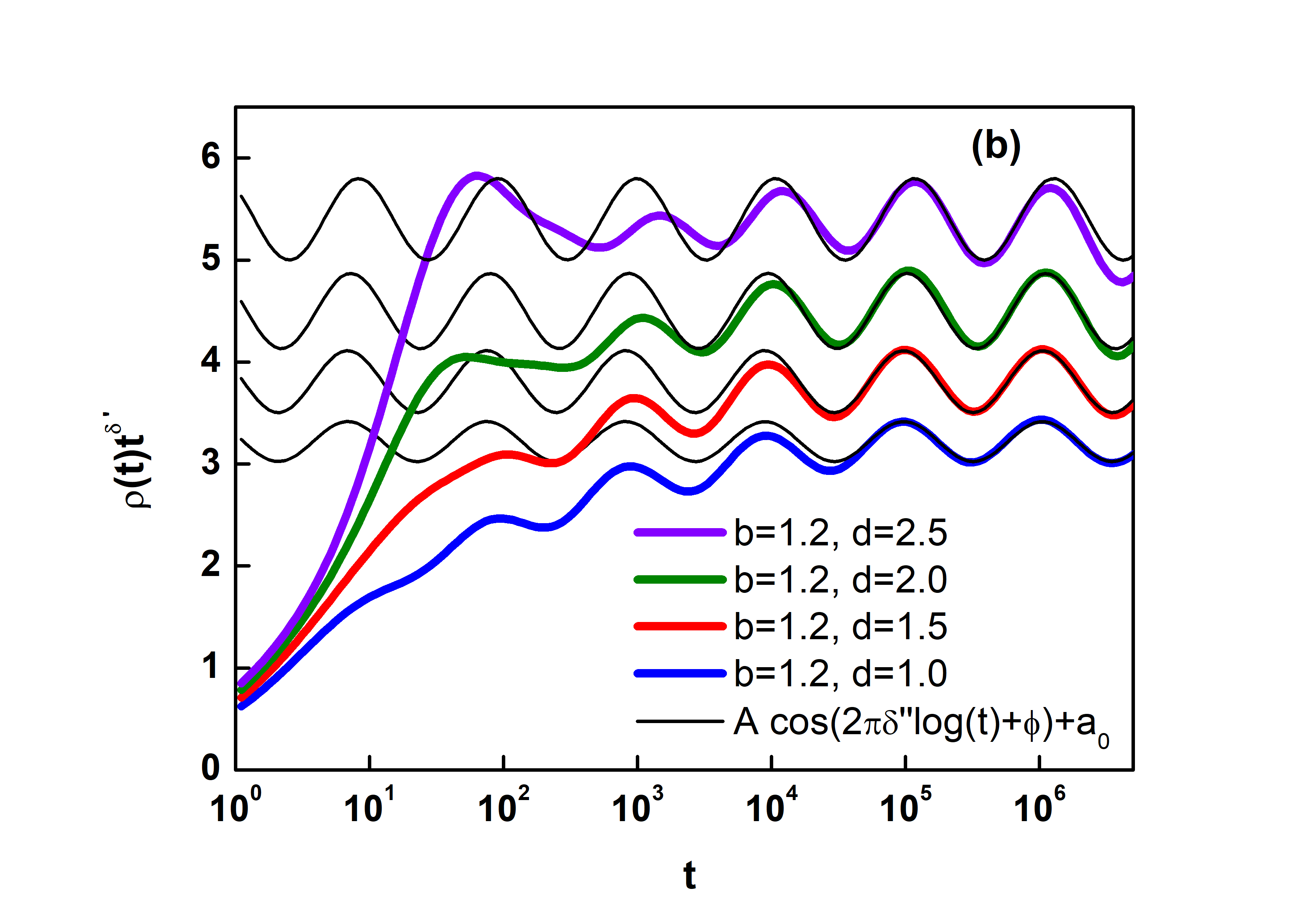}
}
\caption{(a)
shows the plot of $\rho (t)t^{\delta'}$ with time for various values of $b$
(from top to bottom in increasing order of $b$. We assume functional form
$\tau(x)=\exp(bx - d{\frac{\cos(\pi x)}{\sqrt{x}}}) $ for relaxation times.
The log periodic oscillations are evident in the picture.
However $\delta'$ and $\delta''$ keep changing with $b$. 
Onset of oscillations is delayed for higher wavelength oscillations.
(b)shows the plot of $\rho(t)t^{\delta'}$ with time for various values of $d$
(From top to bottom in decreasing order of $d$) while $b=1.2$,
$c=1$,$\delta'=0.765$, $\delta''=0.413$.
Both $\delta'$ and $\delta''$ remain unchanged.
But the onset of logarithmic oscillations is delayed for small $\vert d\vert$.
The y-axis is multiplied by an arbitrary constant for better visualization
and $\omega=2\pi\delta''$
in both figures. 
}
\label{Fig:8}
\end{figure}

\begin{figure}
\scalebox{0.3}{
\includegraphics{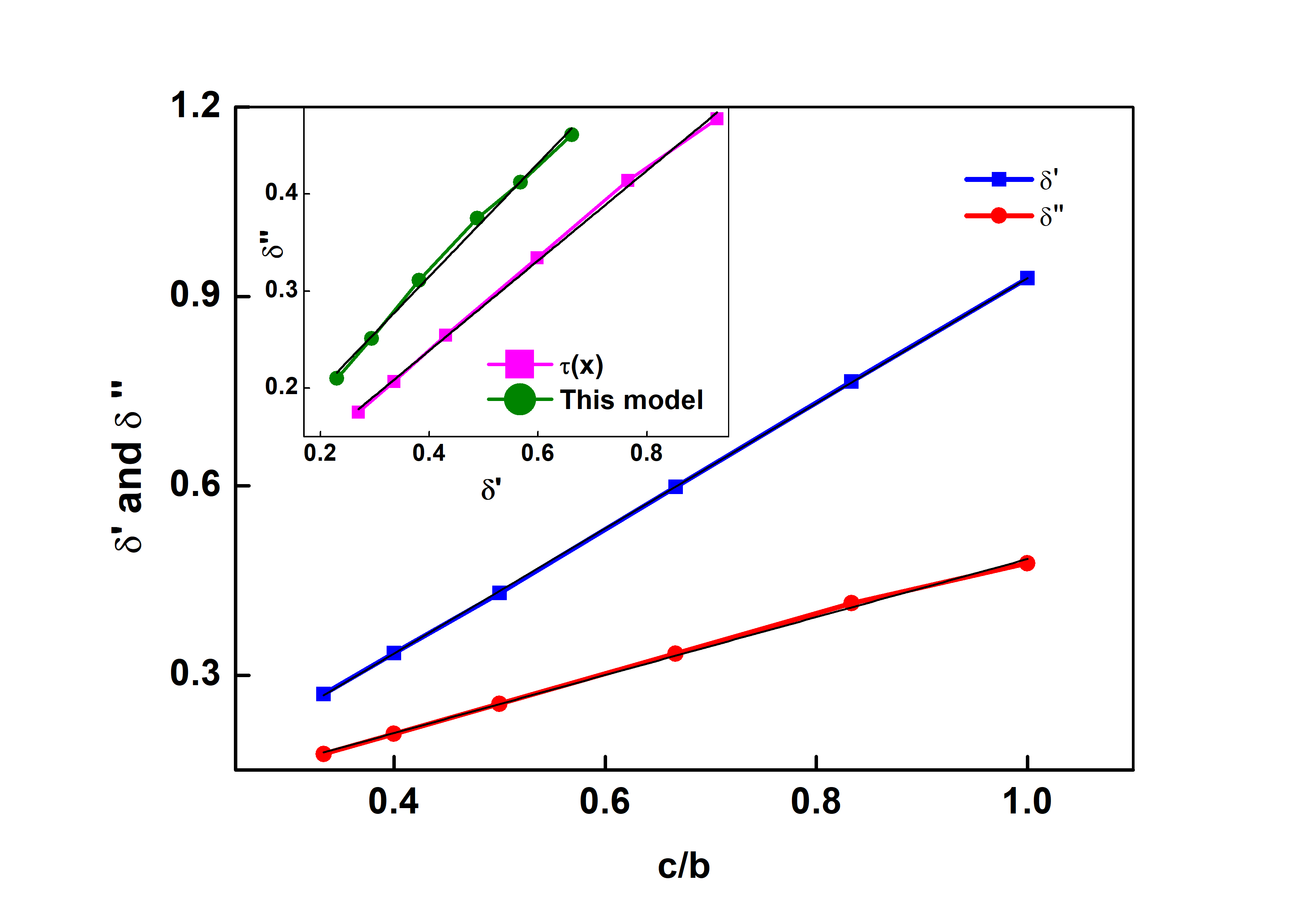}
}
\caption{shows the change in values of $\delta'$ and $\delta''$ with $c/b$ where $c=1$.
The Fig. in inset shows the plot of $\delta'$ and $\delta''$ obtained for
simulation of $\tau(x)$ and 1-d model.
We find that there is linear relationship between two exponents in both cases.}
\label{Fig:9}
\end{figure}

We extend this study to 2-D and 3-D. Let us consider the state of a site at time $t$,
$\textit{v}_(i,j)(t)$ on lattice of size $N^2$.
Let the sites be of four types:
$L$, $R$, $U$ and $D$. For all four types, the evolution occurs according to
the value of the site itself and one of its nearest neighbors (whichever is not
blocked by the quenched defects).
The conditional probabilities 
$P_L(\textit{v}_{i,j}(t+1)|\textit{v}_{i,j}(t)+\textit{v}_{i,j-1}(t))$ and 
$P_R(\textit{v}_{i,j}(t+1)|\textit{v}_{i,j}(t)+\textit{v}_{i,j+1}(t))$ and 
$P_U(\textit{v}_{i,j}(t+1)|\textit{v}_{i,j}(t)+\textit{v}_{i-1,j}(t))$ and
$P_D(\textit{v}_{i,j}(t+1)|\textit{v}_{i,j}(t)+\textit{v}_{i+1,j}(t))$ 
are defined as follows.
$P_L(1\vert k)=P_R(1\vert k)=P_U(1\vert k)=P_D(1\vert k) =p$ for $k\ne 0$ 
and 0 for $k=0$.	
We introduce a defect in the lattice by assigning type $L$, $U$, $R$ and $D$
to each site. This results in the coupling of a given site in a certain
randomly chosen direction. At time $t=0$ almost half of the $N \times N$ sites are
randomly chosen as 'active'. Boundary condition are analogous to 1-d case.
In 2-d, we define boundary condition as follows:
if $i-1 <1,j-1<1$ then $i-1\equiv N+i-1,j-1\equiv N+j-1$ 
and if $i+1 >N, j+1>N$ then $i+1\equiv i+1-N,j+1\equiv j+1-N$
In the case of 3-D, we have six possible
directions. 
Each site is
coupled with neighbors only in one direction
and the conditional probabilities for their
evolution can be defined on the similar lines. 
In all  these  cases $\rho(t) \rightarrow 0$ 
 as $t \rightarrow \infty$ for $p<1$.
 
In the 2-D model, we simulate a 2-D lattice of size $N^2$ where $N=600$ for a
very long time $t=1\times 10^{7}$ and average over more than $600$
configurations. In 3-D we consider a lattice
of size $N^3$ where $N=50$ for time up to $5\times 10^6$
and average over more than $10^3$ configurations.
In both cases,
$\rho(t)$ decays to an absorbing phase for $p\ne 0$.
For small values of $p$, $\rho(t)$ 
decays exponentially or stretched exponentially to an inactive
state.
As in 1-D, 
a regime of continuously
changing power law decay with complex exponent  $\delta=\delta'+i 2\pi \delta''$
is observed for $p$ close to 1. 
The fact that the exponent is complex is reflected
in logarithmic oscillations in the decay of
$\rho(t)$  particularly
as $p\rightarrow 1$  in both 2-D and 3-D(See Fig.\ref{Fig:6}(a)and (b)).

In 2-D and 3-D, we connect each site with only one neighbor.
This neighbor is affected by only one neighbor and so on.
This can be viewed
as a random walk. Thus if we start this walk from a site that does not
affect any other site and continue till the walker
backtracks or gets connected to a site visited previously, the
walk ends there since each site couples to only one site.
A few examples of such walks are shown in Fig.\ref{Fig:7}(a),(b), and (c).
The fraction of such surviving walks goes down exponentially in 2-D as well
as 3-D, though the exponent is smaller in 3-D.(See Fig.\ref{Fig:7}(d))
Now if all the sites covered by this walk go to zero, these sites
will be in an absorbing state forever since all the sites which they are
affected by being in an absorbing state. These sites form a 1-D lattice
of finite size
in all practical senses. This leads to fragmentation of
lattice and the mechanism in higher dimensions could be the same
as in one dimension.

Now the question is what is the origin of the complex exponents?
In the case of the 1-D model, we simulate systems of $k+2$ sites such 
that the first
and the $k+2^{th}$ are set as inactive 
and the rest of the sites are kept active.
Thus we have a group  of $k$ sites of
evolving according to rules of the 1-D model as described in
the above section. 
For uncorrelated disorder, we expect the probability of 
occurance of group of a given size to decay exponentially.
We denote the average time taken by this group
to become inactive by $T_k$. 
We note that the fraction of groups of size $k$ surviving 
till time $t$ seems to be a function of $t/T_k$ (See Fig.5)
Thus $T_k$ is a characteristic time for system size $k$.
This average time $T_k$ increases exponentially 
with relation $T_k \sim \exp(\gamma k)$ with
oscillations imposed over and above the exponential (See Fig\ref{Fig:5}(a)).
(In Fig\ref{Fig:5}(b), we plot $T_k/\exp(\gamma_p k)$ versus $k$
for various values of $p$.)
We argue that the combination of exponentially rare regions which survive for
exponentially long times
leads to a power law and the oscillations over and above
this exponential leads to log-periodicity.

The standard explanation of Griffiths phase goes as follows. 
The probability of finding rare regions is exponentially small $P(x)\propto \exp(-cx)$ but
can exhibit exponentially large lifetime $\tau(x)=\tau_0(x)= \exp(bx)$.
The fraction of active sites  can be approximated 
by $\rho(t) \sim \int$ $x P(x)\exp{(-t/\tau(x))} dx$.
Using saddle-point approximation, it can be shown that this leads
to algebraic decay of $\rho(t)$, {\it{i.e.}} $\rho(t)\sim t^{-c/b}$
with continuously varying exponent.
We do not exactly know the functional form of relaxation times for different sizes.
However, 
in this model, the oscillations get damped, and we propose a functional form
as $\tau(x) \sim \tau_0(x) \tau_p(x)$
where $\tau_p(x)= \exp(-d \frac{\cos(\pi x)}{\sqrt{x}})$ because the odd-even oscillations
get damped quickly. (Sign of $d$ in the
expression of $\tau_p(x)$ does not matter because $\cos$ takes either sign.) We 
numerically compute the above sum
$\rho(t) = \sum_{x=1}^{N}x \exp(-cx) \exp(-t \exp(-bx + d\frac{\cos(\pi x)}{\sqrt{x}}))$ for large $N$, 
{\it{i.e}} $N=10^7$.
The plot of $\rho(t) t^{\delta'}$ $\textit{vs}$ time clearly shows log periodic oscillations as shown in Fig.\ref{Fig:8}(a).
The values of $\delta'$ and $\delta''$ do not depend on $d$ as shown in Fig.\ref{Fig:8}(b). 
 The obtained values of $\delta'$ is close to $c/b$ as expected and it does not depend on 
$d$ at all. Similarly $\delta''$ does not change for any $\vert d \vert \ne 0$.  
However, for smaller values of $\vert d\vert$, 
the amplitude of oscillations is reduced and the onset of oscillations is delayed. 
Both 
exponents vary linearly with $\frac{c}{b}$ and are expected to vary linearly with each other. 
Although $\tau(x)$ is not a perfect analogy of the model presented here, the above simulations show that
the lifetime
of rare region imposed with the periodic term can generate
log periodic oscillations. 
The exponents $\delta'$ and $\delta''$ have a linear relationship in exponents
computed using this ansatz as well as in our model. 
This linear variation is shown in Fig.\ref{Fig:9}.  
The onset of oscillations is delayed when oscillations have
longer periodicity in our model. We observe it in our ansatz as well.  
We have studied a few different functional forms of $\tau_p(x)$ and they lead to log-periodic oscillations
as well. Thus we believe that odd-even oscillations over and above the exponential in relaxation
times are the likely reason for log-periodic oscillations.

{\it{Summary:}}
We have studied a contact process with random asymmetric couplings
in 1-3 dimensions on Domnay-Kinzel automaton. 
In one-dimension, we study a system in which each lattice site is
coupled to two neighbors either on left or on right. In two
and three dimensions
each site is coupled to a neighbor chosen randomly. This is a quenched
disorder.
For low values of $p$ the fraction of active sites
$\rho(t)$ decays exponentially or stretched exponentially.
But for $p\rightarrow 1$,
$\rho(t)$ shows a power law decay
with a complex exponent. Thus
we observe log-periodic oscillations in time over and above the power law decay.
This power law decay of order parameter
with complex exponent can be termed as a complex Griffiths phase.
Such a transition is not observed when the lattice
is not effectively fragmented
in disjoint units. We have also given an argument that this is likely
an effect of odd-even oscillations in relaxation times as a function
of size.

In the complex Griffiths phase,
the real part of the exponent decreases continuously   
while  amplitude and wavelength of oscillations increases 
as $p \rightarrow 1$.
This is a system with a parallel update of all sites.
It can be of interest to study the impact of changes in
updating schemes, dimensionality, and other factors in
the complex Griffiths phase.

{\it{Acknowledgement}}
PMG thanks DST-SERB (EMR/2016/006685) for funding and Professor P. Sen and Professor M. Burma for discussions.
\bibliography{new}

\end{document}